\documentclass[ALICE,manyauthors,nocleardouble]{cernphprep}

\usepackage{amssymb}
\usepackage{amsmath}
\usepackage{mciteplus}
\usepackage{hyperref}
\usepackage{graphicx}
\usepackage{dcolumn}
\usepackage{bm}
\usepackage{lineno}
\usepackage{float}
\usepackage{color}

\usepackage{epstopdf}
\usepackage[normalem]{ulem}

\newcommand{ \mean }[1]{\left\langle #1 \right\rangle}   
\newcommand{\pT}{p_{\rm{T}}}
\newcommand{\pspec}{{\rm p}}
\newcommand{\tspec}{{\rm t}}

\newcommand{\tpspec}{{\rm t,p}}
\newcommand{\Qc}{{\rm Q}}
\newcommand{\Qv}{{\rm \bf Q}}
\newcommand{\Qx}{\Qc_x}
\newcommand{\Qy}{\Qc_y}
\newcommand{\Qxy}{\Qc_{x(y)}}

\newcommand{\sNN}{\sqrt{s_{\rm NN}}}

\newcommand{\Qtp}{\Qv^\tpspec}
\newcommand{\Xp}{\Qx^\tspec}
\newcommand{\Xm}{\Qx^\pspec}
\newcommand{\Yp}{\Qy^\tspec}
\newcommand{\Ym}{\Qy^\pspec}
\newcommand{\Xpm}{\Qx^\tpspec}
\newcommand{\Ypm}{\Qy^\tpspec}

\newcommand{\XYt}{\Qxy^\tspec}
\newcommand{\XYp}{\Qxy^\pspec}

\newcommand{\vodd}{v_1^{\rm odd}}
\newcommand{\veven}{v_1^{\rm even}}
\newcommand{\PsiSPp}{\Psi^\pspec_{\rm SP}}
\newcommand{\PsiSPt}{\Psi^\tspec_{\rm SP}}

\newcommand{\PsiSP}{\Psi_{\rm SP}}
\newcommand{\PsiPP}{\Psi_{\rm PP}^{(1)}}
\newcommand{\PsiRP}{\Psi_{\rm RP}}
\newcommand{\meanpt}{\left<\pT\right>}
\newcommand{\meanpx}{\left<p_{x}\right>}
\newcommand{\pxshift}{\meanpx/\meanpt}
\newcommand{\pxodd}{\meanpx^{\rm odd}}
\newcommand{\pxeven}{\meanpx^{\rm even}}
\newcommand{\pxoddshift}{\pxodd/\meanpt}

\newcommand{\vp}{v_1\{\PsiSPp\}}
\newcommand{\vt}{v_1\{\PsiSPt\}}

\begin{document}

\begin{titlepage}
\PHnumber{2013-100}               
\PHdate{16 June 2013}

\title{Directed flow of charged particles at mid-rapidity relative to the spectator plane 
in Pb--Pb collisions at $\sNN=$~2.76~TeV
}
\ShortTitle{Directed flow of charged particles at mid-rapidity}
\Collaboration{ALICE Collaboration%
\thanks{See Appendix~\ref{app:collab} for the list of collaboration members}}

\ShortAuthor{ALICE Collaboration}

\begin{abstract}
The directed flow of charged particles at mid-rapidity is measured in Pb--Pb collisions 
at $\sNN=$ 2.76~TeV relative to the collision symmetry plane defined by the spectator nucleons. A negative slope of the rapidity-odd directed flow component
with approximately 3 times smaller magnitude than found at the highest RHIC energy is observed. This
suggests a smaller longitudinal tilt of the initial system
and disfavors the strong fireball rotation predicted for the LHC energies.
The rapidity-even directed flow component is measured for the first time with spectators and
found to be independent of pseudorapidity with a sign change at transverse momenta $\pT$ between $1.2$ and $1.7$~GeV/$c$.
Combined with the observation of a
vanishing rapidity-even $\pT$ shift along the spectator deflection
this is strong evidence for dipole-like initial density
fluctuations in the overlap zone of the nuclei.
Similar trends in the rapidity-even directed flow and the estimate
from two-particle correlations at mid-rapidity, which is larger by about a factor of 40,
indicate a weak correlation between fluctuating participant and spectator symmetry planes.
These observations open new possibilities for investigation
of the initial conditions in heavy-ion collisions with spectator nucleons.
\end{abstract}
\end{titlepage}

The goal of the heavy-ion program at the Large Hadron Collider (LHC)
is to explore the properties of deconfined quark-gluon matter.
Anisotropic transverse flow is sensitive to the early 
times of the collision, when the deconfined state of quarks and gluons is expected
to dominate the collision dynamics (see reviews~\cite{Reisdorf:1997fx,Herrmann:1999wu,Voloshin:2008dg} and references therein), with a positive (in-plane) elliptic flow as first observed 
at the Alternating Gradient Synchrotron (AGS)~\cite{Barrette:1994xr,Barrette:1996rs}.
A much stronger flow was subsequently measured at the
Super Proton Synchrotron (SPS)~\cite{Appelshauser:1997dg},
Relativistic Heavy Ion Collider (RHIC)~\cite{Ackermann:2000tr,Adcox:2002ms,Back:2004zg} and
recently at the LHC~\cite{Aamodt:2010pa,Chatrchyan:2012ta,ATLAS:2012at}.
Elliptic flow at RHIC and the LHC is reproduced by hydrodynamic model calculations
with a small value of the ratio of shear viscosity to entropy density~\cite{Huovinen:2001cy,Heinz:2011kt,Song:2007fn,Song:2007ux}.
Despite the success of hydrodynamics in describing
the equilibrium phase of matter produced in a relativistic heavy-ion collision,
there are still large theoretical uncertainties in determination of the initial conditions.
Significant triangular flow measured recently at RHIC~\cite{Adamczyk:2013waa,Adare:2011tg} and LHC~\cite{ALICE:2011ab,Chatrchyan:2012wg,ATLAS:2012at}
energies has demonstrated~\cite{Alver:2010gr,Sorensen:2010zq}
that initial energy fluctuations play an important role in the development of the
final momentum-space anisotropy of the distribution of produced particles.

\begin{figure}[ht]
\begin{center}
\includegraphics[width=.49\textwidth]{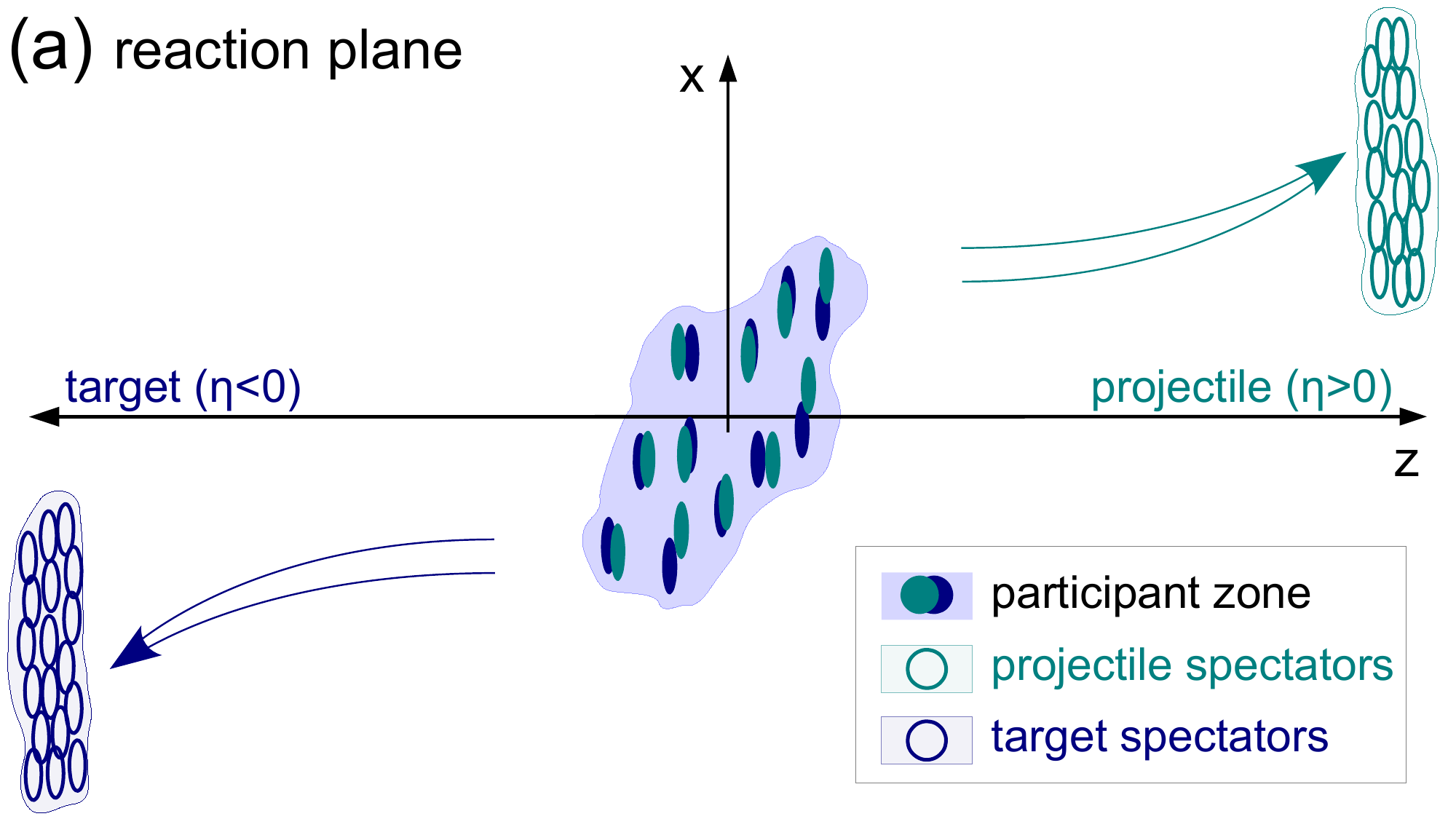}~~%
\includegraphics[width=.49\textwidth]{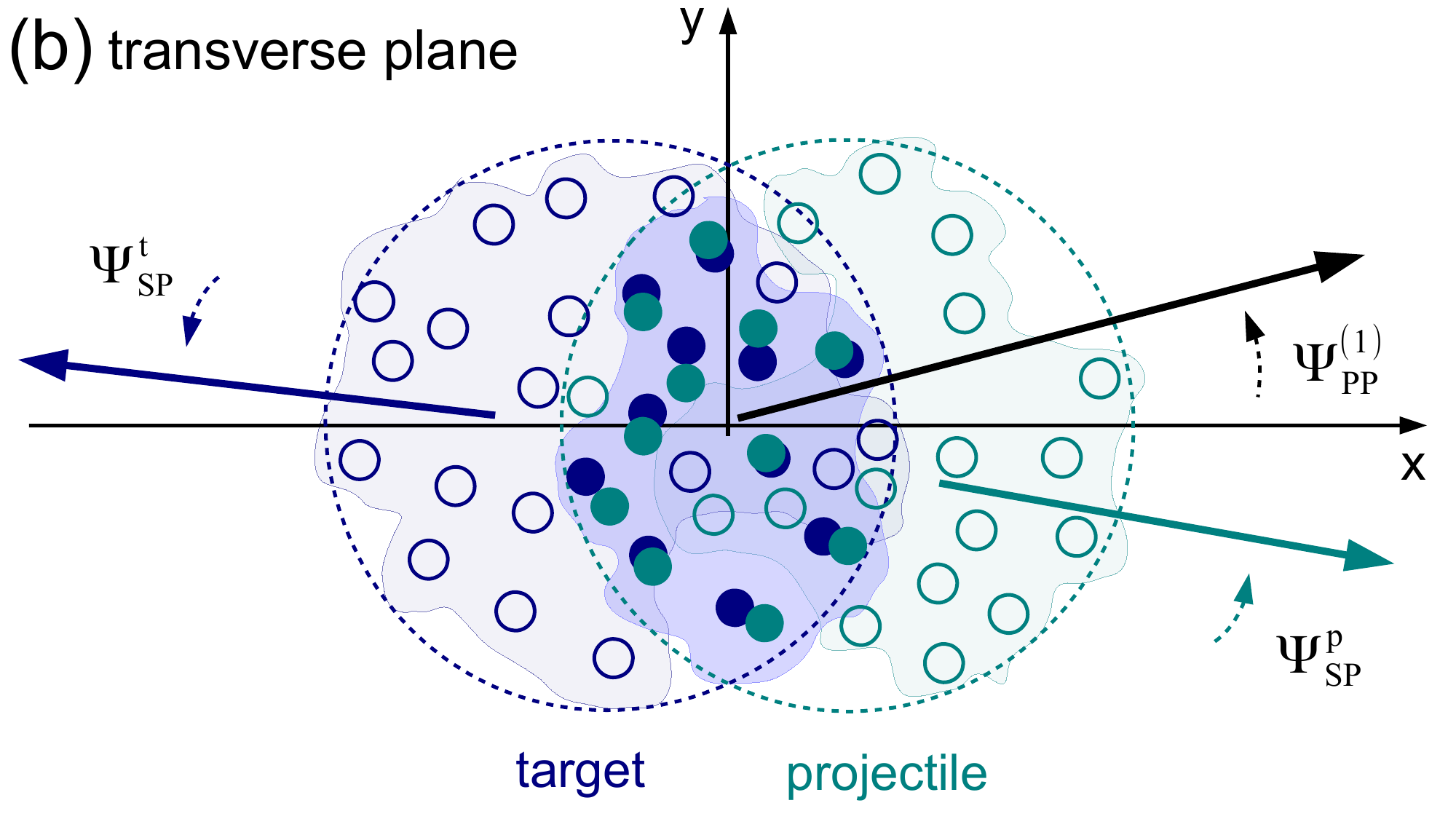}
\caption{(color online)
Sketch of a non-central heavy-ion collision. See text for description of the figure.
}
\label{fig:1}
\end{center}
\end{figure}
The collision geometry is illustrated in Fig.~\ref{fig:1}
which depicts the participant overlap region and spectators as viewed in
(a) the reaction plane and (b) the plane perpendicular to the beam.
Figure~\ref{fig:1}(a) shows the projectile and target spectators
repelled in the reaction ($xz$) plane from the center of the colliding system along the impact parameter ($x$) direction.
An alternative scenario where spectators are attracted towards the center of the system is discussed in~\cite{Alvioli:2010yk}.

The directed flow is characterized by the first harmonic coefficient $v_1$
in a Fourier decomposition of the particle azimuthal distribution
with respect to one of the collision symmetry planes, $\Psi$,
as illustrated in Fig.~\ref{fig:1}(b) and discussed below
\begin{equation}
\label{Eq:v1}
v_1(\eta,\pT)\{\rm \Psi\}= \left< \cos(\phi - \Psi) \right>.
\end{equation}
Here $\eta=-\ln\left[\tan\left(\theta/2\right)\right]$,
$\pT$, $\theta$ and $\phi$ are the particle
pseudorapidity, transverse momentum, polar and azimuthal angles, respectively.
The brackets "$\left<...\right>$" indicate an average
over measured particles in all recorded events.

For a non-fluctuating nuclear matter distribution, the directed flow in the
participant zone develops along the impact parameter direction.
The collision symmetry requires that the directed flow be
an anti-symmetric function of pseudorapidity, \mbox{$\vodd(\eta) = - \vodd(-\eta)$}.
Due to event-by-event fluctuations
in the initial energy density of the collision,
the participant plane angle ($\PsiPP$) defined by the dipole asymmetry
of the initial energy density~\cite{Teaney:2010vd,Luzum:2010fb}
and that of projectile ($\PsiSPp$) and target ($\PsiSPt$) spectators are different from the geometrical reaction plane angle $\PsiRP$ ($x$-axis in Fig.~\ref{fig:1}(b)).
As a consequence, the directed flow can develop~\cite{Teaney:2010vd,Luzum:2010fb,Gardim:2011qn,Retinskaya:2012ky}
a rapidity-symmetric component, $\veven(\eta) = \veven(-\eta)$,
which does not vanish at mid-rapidity.

The slope of $\vodd$ as a function of rapidity
at AGS~\cite{Barrette:1996rs,Barrette:1997pt} and SPS~\cite{Alt:2003ab,Aggarwal:1998vg}
energies  is driven by the difference between baryon and meson production and shadowing by the nuclear remnants.
At higher (RHIC) energies a multiple zero crossing
of $\vodd$ with rapidity outside the nuclear fragmentation regions
was predicted as a signature of the deconfined phase transition~\cite{Brachmann:1999xt,Csernai:1999nf}.
However, the RHIC measurements~\cite{Adams:2005ca,Abelev:2008jga,Back:2005pc,Agakishiev:2011id} did not reveal such a structure.
The magnitude of the directed flow depends on
the amount of baryon stopping in the nuclear overlap zone~\cite{Snellings:1999bt}. The
 two can be related via realistic model calculations, making $\vodd$ an important experimental probe
of the initial conditions in a heavy-ion collision.
The initial conditions assumed in model calculations of
$\vodd$ range from incomplete baryon stopping~\cite{Snellings:1999bt},
with a positive space-momentum correlation,
to full nucleon stopping with a tilted~\cite{Csernai:1999nf,Bozek:2010bi} or
rotating~\cite{Csernai:2011gg} source of matter produced in the overlap zone of the nuclei.
Model calculations generally agree on
the negative sign of the $\vodd$ slope as a function of pseudorapidity at
RHIC~\cite{Adams:2005ca,Abelev:2008jga,Back:2005pc,Agakishiev:2011id}.
The model predictions for $\vodd$ at the LHC
vary from having the same slope as at RHIC but with smaller magnitude~\cite{Bozek:2010bi}
to an opposite (positive) slope with significantly
larger magnitude~\cite{Bleibel:2007se,Csernai:2011gg}.

The $\veven$ estimated from the two-particle azimuthal correlations at mid-rapidity
for RHIC~\cite{Agakishiev:2010ur} (see also \cite{Luzum:2010fb})
and LHC~\cite{Aamodt:2011by,ATLAS:2012at,Chatrchyan:2012wg} energies
is in approximate agreement with ideal hydrodynamic model
calculations~\cite{Gardim:2011qn,Retinskaya:2012ky}
for dipole-like~\cite{Teaney:2010vd} energy fluctuations
in the overlap zone of the nuclei.
Interpretation of the two-particle correlations is complicated due to
a possibly large bias from correlations unrelated to the initial geometry (non-flow)
and due to the model dependence of the correction procedure
for effects of momentum conservation~\cite{Retinskaya:2012ky}.
The directed flow measured relative to the spectator deflection
is free from such biases and provides a cleaner
probe of the initial conditions in a heavy-ion collision.
It also allows for a study of the main features of the dipole-like energy
fluctuations such as a vanishing transverse momentum shift
of the created system along the direction of the spectator deflection.
Directed flow and its fluctuations also play an important role in understanding
effects due to the strong magnetic field in heavy-ion collisions~\cite{Teaney:2010vd} and interpretation
of the observed charge separation relative to the reaction plane~\cite{Abelev:2012pa}
in terms of the chiral magnetic effect~\cite{Fukushima:2008xe}.

In this Letter, we report the
charged particle directed flow measured relative to the deflection of spectator neutrons
in Pb--Pb collisions at $\sNN=$~2.76~TeV. About 13~million minimum-bias~\cite{Aamodt:2010pa} collisions in the \mbox{5-80\%} centrality range were analyzed.
For the most central \mbox{(0-5\%)} collisions, the small number of spectators does
not allow for a reliable reconstruction of their deflection. Two forward scintillator arrays (VZERO)~\cite{Aamodt:2008zz} were used to determine the collision centrality.
Charged particles reconstructed in the Time Projection Chamber (TPC)~\cite{Alme:2010ke}
with $\pT>0.15$~GeV/$c$ and $|\eta|<0.8$
were selected for the analysis.

The deflection of neutron spectators was reconstructed
using a pair of Zero Degree Calorimeters (ZDC)~\cite{Carminati:2004fp}
with $2\times 2$ segmentation
installed 114 meters from the interaction point on each side,
covering the $|\eta| > 8.78$ (beam rapidity) region.
A typical energy measured by both ZDCs for 30-40\% centrality is about 100 TeV~\cite{Abelev:2013qoq}.
The spectator deflection in the transverse plane was measured with a pair of two-dimensional vectors
\begin{equation}
\Qtp \equiv \left(\Xpm,\Ypm\right) =  \left.\sum\limits_{i=1}^{4}{\bf n}_i E_i^\tpspec\right/{\sum\limits_{i=1}^{4}} E_{i}^\tpspec,
\label{Eq:Qvector}
\end{equation}
where ${\rm p}$ (${\rm t}$) denotes the ZDC on the $\eta >0$ ($\eta<0$) side of the interaction point,
$E_i$ is the measured signal and ${\bf n}_i=\left(x_i,y_i\right)$ are the coordinates of the $i$-th ZDC segment.
An asymmetry of 0.1\%~\cite{ALICE:2012aa} in energy calibration of the two ZDCs 
and an absolute energy scale uncertainty cancel in Eq.~(\ref{Eq:Qvector}).
To compensate for the run-dependent variation of the LHC beam crossing position,
an event-by-event correction $\Qtp \to \Qtp-\mean{\Qtp}$~\cite{Voloshin:2008dg}
was applied as a function of collision centrality
and transverse position of the collision vertex relative to the center of the ALICE detector.
Experimental values of the correction for the 30-40\% centrality class
are $\left<\right.\XYp\left.\right>\approx2.0~(-1.5)$~mm and \mbox{$\left<\right.\XYt\left.\right>\approx-1.1~(0.01)$~mm}.

The directed flow is determined with the scalar product
method~\cite{Voloshin:2008dg,Selyuzhenkov:2007zi}
\begin{eqnarray}
\nonumber
\vp=\frac{1}{\sqrt{2}}\left[
\frac{{\langle u_x \Xm \rangle}}{{\sqrt{\left|\langle \Xp \Xm\rangle\right|}}}
+
\frac{{\langle u_y \Ym \rangle}}{{\sqrt{\left|\langle \Yp \Ym\rangle\right|}}}
\right],~\\
\vt = -
\frac{1}{\sqrt{2}}\left[
\frac{{\langle u_x \Xp \rangle}}{{\sqrt{\left|\langle \Xp \Xm\rangle\right|}}}
+
\frac{{\langle u_y \Yp \rangle}}{{\sqrt{\left|\langle \Yp \Ym\rangle\right|}}}
\right]
\label{Eq:v1SP}
\end{eqnarray}
where $u_x=\cos\phi$ and $u_y=\sin\phi$ are defined for charged particles at midrapidity.
The odd and even components of the directed flow relative to the spectator plane
($\Psi=\PsiSP$ in Eq.~(\ref{Eq:v1})) are then calculated from the equations
\begin{equation}
\vodd\{\PsiSP\} = \left.\left[\vp+\vt\right]\right/2
\label{Eq:v1odd}
\end{equation}
and
\begin{equation}
\veven\{\PsiSP\} = \left.\left[\vp-\vt\right]\right/2.
\label{Eq:v1even}
\end{equation}
Equation~(\ref{Eq:v1odd}) defines the sign
of $\vodd$ using the convention
used at RHIC~\cite{Adams:2005ca,Abelev:2008jga} and
implies a positive directed flow (or deflection along the positive $x$-axis
direction in Fig.~\ref{fig:1}(a)) of the projectile spectators.

The negative correlations
$\mean{\Xp \Xm}$ and $\mean{\Yp\Ym}$~\cite{Ref:hepData_ZDCresolution}
indicate a deflection of the projectile and target spectators in opposite directions.
These correlations are sensitive to a combination of the spectator's directed flow relative to the reaction plane
$\PsiRP$ and an additional contribution due to flow of spectators along the fluctuating $\PsiSPp$ and $\PsiSPt$ directions (see Fig.~\ref{fig:1}(b)).
The two contributions are not separable using current experimental techniques and both should be considered in theoretical interpretations of the results derived from Eqs.~(\ref{Eq:v1SP})-(\ref{Eq:v1even}).
Given that the transverse deflection of spectators $\left(d_{\rm spec}\approx \sqrt{\left(\mean{\Xp \Xm}+\mean{\Yp\Ym}\right)/2}\right)$
is tiny compared to the ZDC detector position $|z_{\rm ZDC}|=114$~m along the beam direction,
one can make a rough estimate of the corresponding transverse momentum carried by an individual spectator:
$\pT^{\rm spec} \approx \sNN \left(d_{\rm spec}/z_{\rm ZDC}\right)$.
The measured $d_{\rm spec}$ is about 0.67~(0.92)~mm~\cite{Ref:hepData_ZDCresolution} for the 5-10\% (30-40\%) centrality class
which yields $\pT^{\rm spec}\sim 16~(22)$~MeV/$c$.
Correlations $\langle \Xp \Ym\rangle$ and $\langle \Yp \Xm\rangle$ in orthogonal directions,
which can be non-zero due to residual detector effects,
are less than
5\%~\cite{Ref:hepData_ZDCresolution} of those in the aligned directions.
The 10-20\%~\cite{Ref:hepData_ZDCresolution} difference between $\langle \Xp \Xm\rangle$ and $\langle \Yp \Ym\rangle$
for mid-central collisions is mainly due to a different offset of the beam spot
from the center of the ZDCs in-plane and perpendicular to the LHC accelerator
ring. The corresponding dominant systematic uncertainty is evaluated from the spread of results for different terms in Eq.~(\ref{Eq:v1SP})
and estimated to be below 20\%.
The results obtained with Eq.~(\ref{Eq:v1SP}) are consistent with calculations using the event plane method~\cite{Voloshin:2008dg}.
The results with opposite polarity of the magnetic field of the ALICE detector are consistent within 5\%.
Variation of the results with the collision centrality estimated with the TPC, VZERO,
and Silicon Pixel Detectors~\cite{Carminati:2004fp} and with narrowing the nominal $\pm$10~cm
range of the collision vertex along the beam direction
from the center of the ALICE detector to $\pm$7~cm is less than 5\%.
Altering the selection criteria for the tracks reconstructed with the TPC
resulted in a 3-5\% variation of the results.
The systematic error evaluated for each of the sources
listed above were added in quadrature to obtain the
total systematic uncertainty of the measurement.

\begin{figure}[ht]
\begin{center}
\includegraphics[width=.5\textwidth]{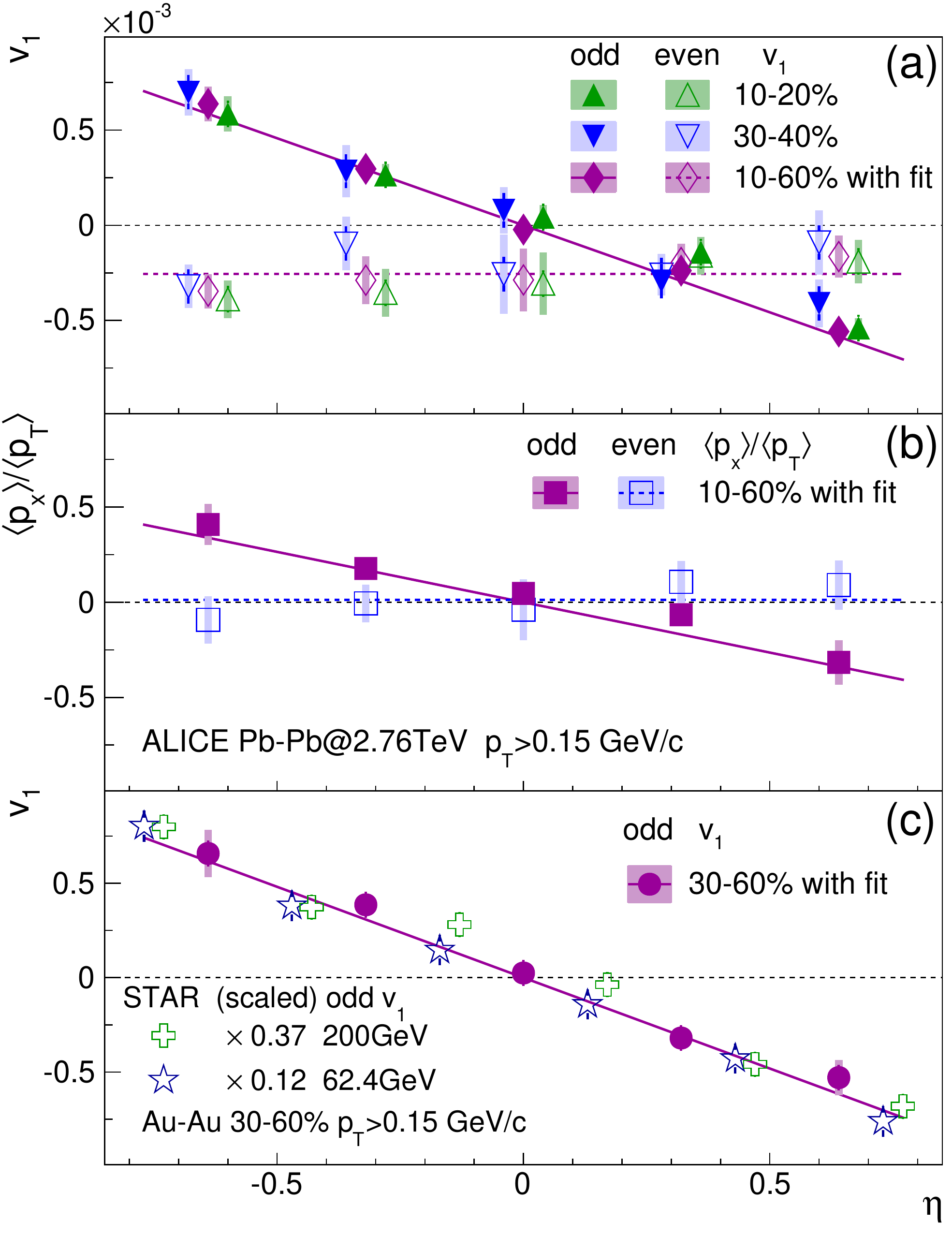}
\caption{(color online)
(a) $v_1$ and (b) $\pxshift$ versus pseudorapidity
in Pb--Pb collisions at $\sNN=$~2.76~TeV.
(c)~$\vodd$ compared to the STAR data~\cite{Abelev:2008jga}
for Au-Au collisions at $\sNN=$~200~(62.4)~GeV downscaled by a factor 0.37~(0.12).
The statistical (systematic) uncertainties are indicated by the error bars (shaded bands).
Lines (to guide the eye) represent fits with a linear (constant) function for $\vodd$ ($\veven$).
}
\label{fig:2}
\end{center}
\end{figure}
Figure~\ref{fig:2}(a) shows the charged particle directed flow as a function of pseudorapidity for 10-20\%, 30-40\%, and 10-60\% centrality classes.
The $\veven(\eta)$ component is found to be negative and independent of $\eta$.
The $\vodd(\eta)$ component exhibits a negative slope as a function of pseudorapidity.
This is in contrast to the positive slope
expected from the model calculations~\cite{Bleibel:2007se,Csernai:2011gg}
with stronger rotation of the participant zone at the LHC than at RHIC. The $\vodd(\eta)$ at the highest RHIC energy~\cite{Abelev:2008jga} has the same sign of the slope and a factor of three larger magnitude. This is consistent with a smaller
tilt of the participant zone in the $x$--$z$ plane
(see Fig.~\ref{fig:1}(a)) as predicted in~\cite{Bozek:2010bi} for LHC energies.
Figure~\ref{fig:2}(c) compares $\vodd$ with the STAR data~\cite{Abelev:2008jga} for
Au-Au collisions at \mbox{$\sNN=200$}~(62)~GeV
downscaled by the ratio 0.37~(0.12) of the slope at the LHC to that at RHIC energy. These ratios indicate 
a strong violation by a factor of 1.82 (4.55) of the beam rapidity scaling  discussed in~\cite{Agakishiev:2011id}.

Figure~\ref{fig:2}(b) shows the relative momentum shift
$\pxshift\equiv\mean{\pT \cos\left(\phi-\PsiSP\right)}/\mean{\pT}$, along the spectator plane
as a function of pseudorapidity.
It is obtained by introducing a $\pT/\mean{\pT}$ weight in front of $u_x$ and $u_y$ in Eq.~(\ref{Eq:v1SP}).
The non-zero $\pxoddshift$ shift has a smaller magnitude than $\vodd$.
The $\pxeven$ vanishes which is consistent with the
dipole-like event-by-event fluctuations of the initial energy density
in a system with zero net transverse momentum.
Disappearance of $\meanpx$ at $\eta\approx0$ indicates
that particles produced at mid-rapidity are not involved
in balancing the transverse momentum carried away by spectators.

\begin{figure}[ht]
\begin{center}
\includegraphics[width=.5\textwidth]{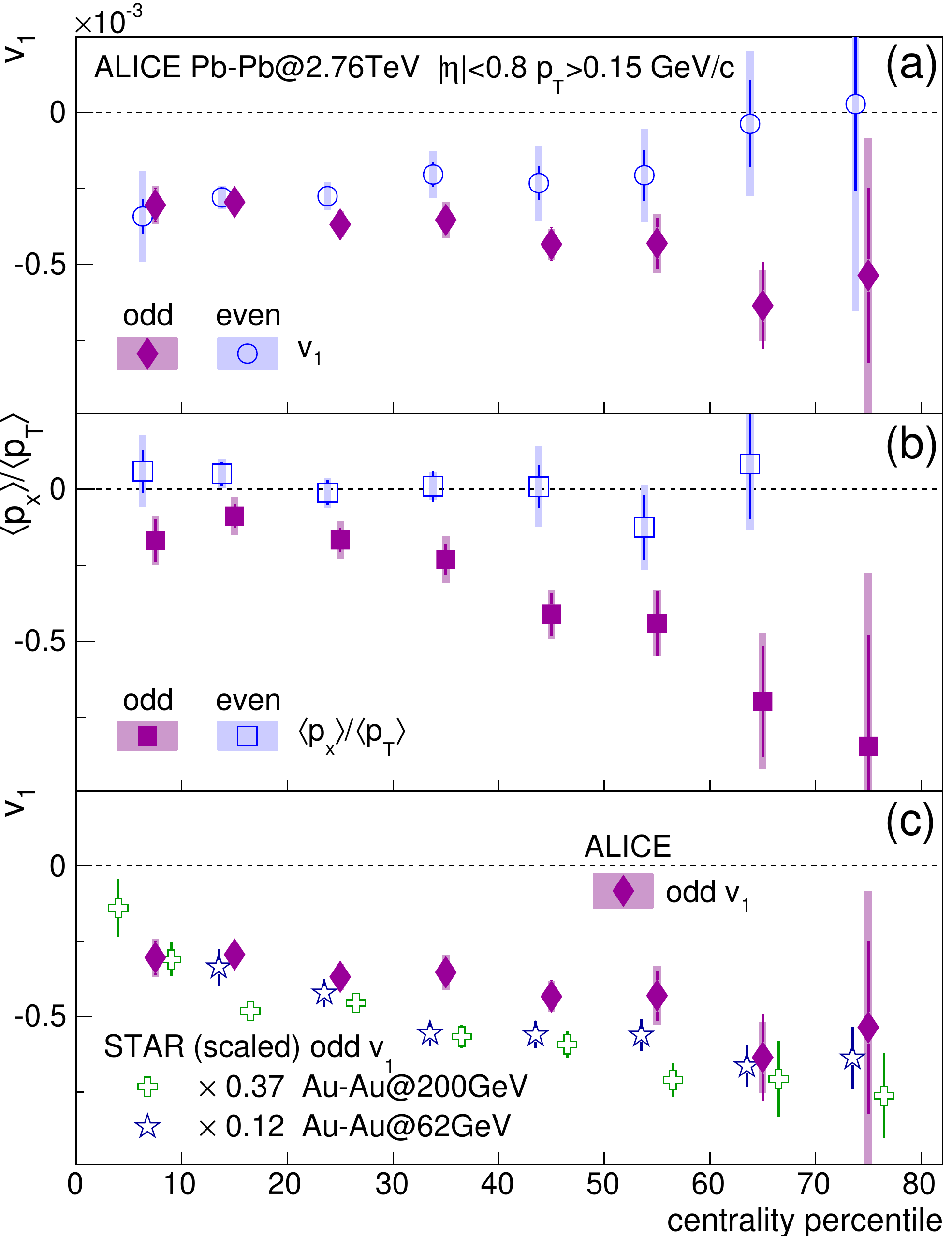}
\caption{(color online)
(a) $v_1$ and (b) $\pxshift$ versus centrality. (c) $\vodd$ comparison with STAR data~\cite{Abelev:2008jga}. See text and Fig.~\ref{fig:2} for description of the data points.
}
\label{fig:3}
\end{center}
\end{figure}
Figures~\ref{fig:3}(a) and \ref{fig:3}(b) present $v_1$ and $\pxshift$ versus collision centrality. The odd components were
calculated by taking values at negative $\eta$ with an opposite sign.
Both $v_1$ components have weak centrality dependence.
The $\pxeven$ component is zero at all centralities, while
$\pxoddshift$ is a steeper function of centrality than $\vodd$.
This suggests that $\vodd$ has two contributions.
The first contribution has a similar
origin as $\veven$ due to asymmetric dipole-like initial energy fluctuations.
The second contribution grows almost linearly from central to peripheral collisions
and represents an effect of sideward collective motion of particles at non-zero rapidity
due to expansion of the initially tilted source. This $\meanpx$ is balanced by that of
the particles produced at opposite rapidity and in very forward (spectator) regions.
The magnitude of $\vodd$ at the LHC is significantly smaller than at RHIC with a similar centrality dependence (see Fig.~\ref{fig:3}(c)).

\begin{figure}[ht]
\begin{center}
\includegraphics[width=.5\textwidth]{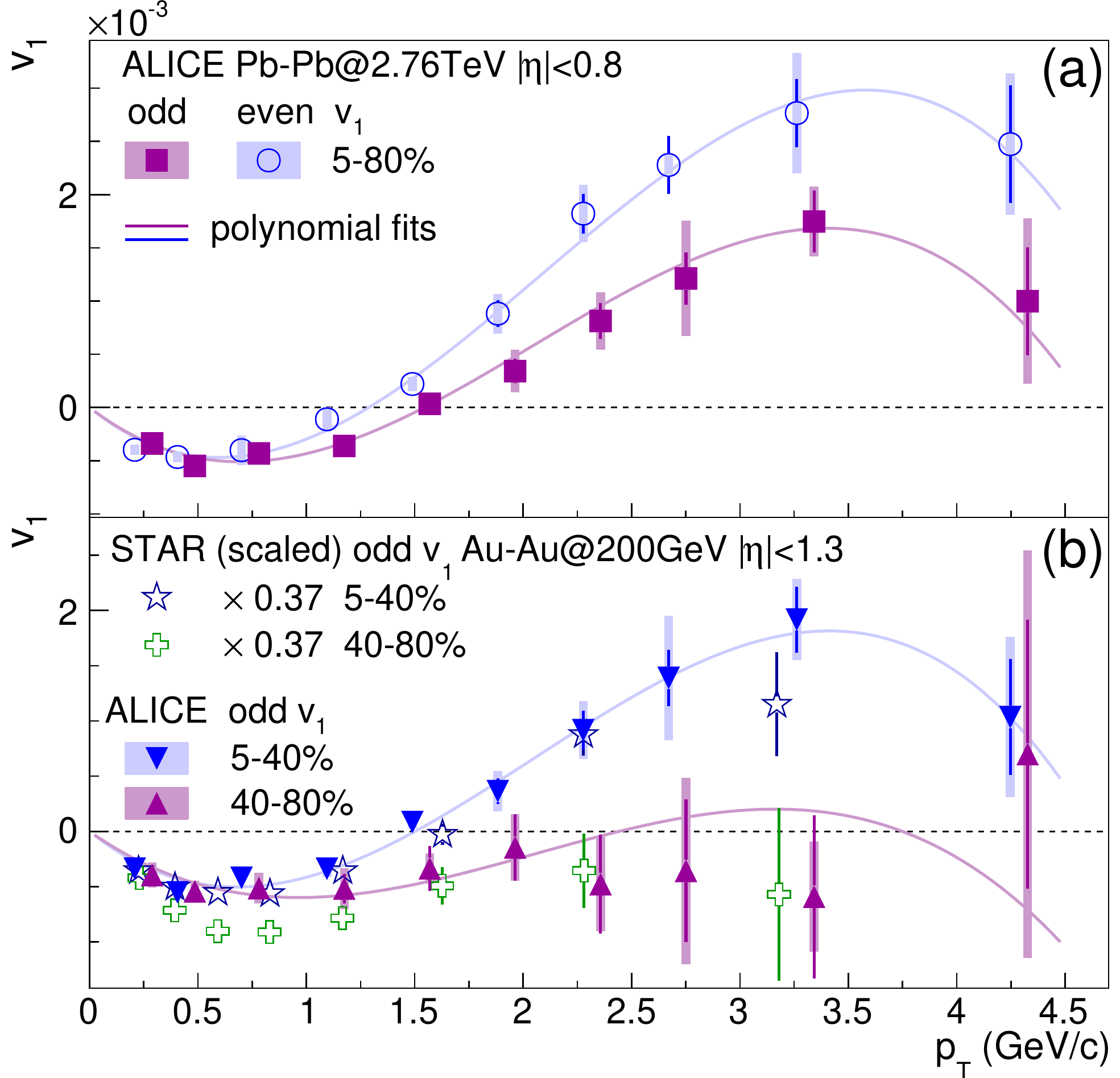}
\caption{(color online)
(a) $v_1$ versus transverse momentum. (b) $\vodd$ comparison with STAR data~\cite{Abelev:2008jga}. See text and Fig.~\ref{fig:2} for description of the data points.
Lines (to guide the eye) represent fits with a third order polynomial.
}
\label{fig:4}
\end{center}
\end{figure}
Figure \ref{fig:4}(a) presents $v_1$ as a function of $\pT$.
Both components change sign around \mbox{$\pT$ between $1.2-1.7$~GeV/$c$} which
is expected for the dipole-like energy fluctuations
when the momentum of the low $\pT$ particles is balanced by those at high
$\pT$~\cite{Teaney:2010vd,Luzum:2010fb,Gardim:2011qn,Retinskaya:2012ky}.
The $\pT$ dependence of $\veven$ relative to $\PsiSP$ is similar to that of $\veven$ relative to $\PsiPP$ estimated from the Fourier fits of the
two-particle correlations~\cite{Aamodt:2011by,ATLAS:2012at,Chatrchyan:2012wg},
while its magnitude is smaller by a factor of
forty~\cite{Retinskaya:2012ky,Selyuzhenkov:2011zj}.
This can be interpreted as a weak correlation,
$\left<\right.\cos(\PsiPP-\PsiSP)\left.\right>\ll~1$,
between the orientation of the participant and spectator collision symmetry planes.
Compared to the RHIC measurements in Fig.~\ref{fig:4}(b),
$\vodd$ shows a similar trend including
the sign change around $\pT$ of $1.5$ GeV/$c$ in central collisions
and a negative value at all $\pT$ for peripheral collisions.

According to hydrodynamic model calculations \cite{Heinz:2004et,Teaney:2010vd,Retinskaya:2012ky}
particles with low $\pT$ should flow
in the direction opposite to the largest density gradient.
This, together with the negative even and odd $v_1$ components relative to $\PsiSP$ measured for particles at mid-rapidity with low transverse momentum
($\pT\lesssim 1.2$~GeV/$c$) allows to determine if spectators deflect away from or towards the center of the system.
However, a detailed theoretical calculation of the correlation between fluctuations in the
spectator positions and energy density in the participant zone such as in \cite{Alvioli:2010yk} is required
to provide a definitive answer to this question.

In summary, the $\vodd$ and $\veven$ components of charged particle directed flow
at mid-rapidity, $|\eta| < 0.8$,  are measured relative to the spectator plane
for Pb--Pb collisions at $\sNN=$~2.76~TeV.
The $\vodd$ has a negative slope as a function of pseudorapidity
with a magnitude about three times smaller than at the highest RHIC energy.
This suggests a smaller tilt of the medium created in the participant zone at the LHC,
with insufficient rotation to alter the slope of
$\vodd(\eta)$ as predicted in~\cite{Bleibel:2007se,Csernai:2011gg}.
As a function of $\pT$, $\vodd$ and $\veven$ cross zero at $\pT$ between $1.2-1.7$~GeV/$c$
for semi-central collisions.
Disappearance of $\meanpx$ for particles produced close to zero rapidity suggest that
they do not play a role in balancing the $\pT$ kick of spectators.
The shape of $\veven(\pT)$ and a vanishing $\pxeven$ is consistent
with dipole-like fluctuations of the initial energy density in the participant zone.
A similar shape but with about forty times larger magnitude
was observed for an estimate of $\veven(\pT)$ relative to the participant plane
from the Fourier fits of the two-particle correlation~\cite{Aamodt:2011by,ATLAS:2012at}.
This indicates that fluctuating participant and spectator collision symmetry planes
are weakly correlated, which is important experimental input for
modeling the ill-constrained initial conditions of a heavy-ion collision.
Future studies of the directed flow at mid-rapidity using identified particles and extension of
the $v_1$ measurements to forward rapidities should provide a stronger
constraint on the effects of initial density fluctuations in the formation of directed flow.

\newenvironment{acknowledgement}{\relax}{\relax}
\begin{acknowledgement}
\section*{Acknowledgements}

The ALICE collaboration would like to thank all its engineers and technicians for their invaluable contributions to the construction of the experiment and the CERN accelerator teams for the outstanding performance of the LHC complex.
%\\
The ALICE collaboration acknowledges the following funding agencies for their support in building and
running the ALICE detector:
 %\\
State Committee of Science,  World Federation of Scientists (WFS)
and Swiss Fonds Kidagan, Armenia,
 %\\
Conselho Nacional de Desenvolvimento Cient\'{\i}fico e Tecnol\'{o}gico (CNPq), Financiadora de Estudos e Projetos (FINEP),
Funda\c{c}\~{a}o de Amparo \`{a} Pesquisa do Estado de S\~{a}o Paulo (FAPESP);
 %\\
National Natural Science Foundation of China (NSFC), the Chinese Ministry of Education (CMOE)
and the Ministry of Science and Technology of China (MSTC);
 %\\
Ministry of Education and Youth of the Czech Republic;
 %\\
Danish Natural Science Research Council, the Carlsberg Foundation and the Danish National Research Foundation;
 %\\
The European Research Council under the European Community's Seventh Framework Programme;
 %\\
Helsinki Institute of Physics and the Academy of Finland;
 %\\
French CNRS-IN2P3, the `Region Pays de Loire', `Region Alsace', `Region Auvergne' and CEA, France;
 %\\
German BMBF and the Helmholtz Association;
%\\
General Secretariat for Research and Technology, Ministry of
Development, Greece;
%\\
Hungarian OTKA and National Office for Research and Technology (NKTH);
 %\\
Department of Atomic Energy and Department of Science and Technology of the Government of India;
 %\\
Istituto Nazionale di Fisica Nucleare (INFN) and Centro Fermi -
Museo Storico della Fisica e Centro Studi e Ricerche "Enrico
Fermi", Italy;
 %\\
MEXT Grant-in-Aid for Specially Promoted Research, Ja\-pan;
 %\\
Joint Institute for Nuclear Research, Dubna;
 %\\
%Korea Foundation for International Cooperation of Science and Technology (KICOS);
National Research Foundation of Korea (NRF);
 %\\
CONACYT, DGAPA, M\'{e}xico, ALFA-EC and the EPLANET Program
(European Particle Physics Latin American Network)
 %\\
Stichting voor Fundamenteel Onderzoek der Materie (FOM) and the Nederlandse Organisatie voor Wetenschappelijk Onderzoek (NWO), Netherlands;
 %\\
Research Council of Norway (NFR);
 %\\
Polish Ministry of Science and Higher Education;
 %\\
National Authority for Scientific Research - NASR (Autoritatea Na\c{t}ional\u{a} pentru Cercetare \c{S}tiin\c{t}ific\u{a} - ANCS);
 %\\
Ministry of Education and Science of Russian Federation, Russian
Academy of Sciences, Russian Federal Agency of Atomic Energy,
Russian Federal Agency for Science and Innovations and The Russian
Foundation for Basic Research;
 %\\
Ministry of Education of Slovakia;
 %\\
Department of Science and Technology, South Africa;
 %\\
CIEMAT, EELA, Ministerio de Econom\'{i}a y Competitividad (MINECO) of Spain, Xunta de Galicia (Conseller\'{\i}a de Educaci\'{o}n),
CEA\-DEN, Cubaenerg\'{\i}a, Cuba, and IAEA (International Atomic Energy Agency);
 %\\
Swedish Research Council (VR) and Knut $\&$ Alice Wallenberg
Foundation (KAW);
 %\\
Ukraine Ministry of Education and Science;
 %\\
United Kingdom Science and Technology Facilities Council (STFC);
 %\\
The United States Department of Energy, the United States National
Science Foundation, the State of Texas, and the State of Ohio.

\end{acknowledgement}

\bibliographystyle{href-physrev4}
\bibliography{paper_v1_midrapidity_spectators}

\newpage
\appendix
\section{The ALICE Collaboration}
\label{app:collab}

% Collaboration: CERN-LHC-ALICE
% Generation Date is 9 Jun 2013 23:00:25 GMT
\begingroup
\small
\begin{flushleft}
B.~Abelev\Irefn{org1234}\And
J.~Adam\Irefn{org1274}\And
D.~Adamov\'{a}\Irefn{org1283}\And
A.M.~Adare\Irefn{org1260}\And
M.M.~Aggarwal\Irefn{org1157}\And
G.~Aglieri~Rinella\Irefn{org1192}\And
M.~Agnello\Irefn{org1313}\textsuperscript{,}\Irefn{org1017688}\And
A.G.~Agocs\Irefn{org1143}\And
A.~Agostinelli\Irefn{org1132}\And
Z.~Ahammed\Irefn{org1225}
% ...
\renewcommand\labelenumi{\textsuperscript{\theenumi}~}

\section*{Affiliation notes}
\renewcommand\theenumi{\roman{enumi}}
\begin{Authlist}
\item \Adef{0}Deceased
\item \Adef{idp4188160}{Also at: M.V.Lomonosov Moscow State University, D.V.Skobeltsyn Institute of Nuclear Physics, Moscow, Russia}
\item \Adef{idp4480544}{Also at: University of Belgrade, Faculty of Physics and "Vin\v{c}a" Institute of Nuclear Sciences, Belgrade, Serbia}
\item \Adef{idp5499968}{Also at: Institute of Theoretical Physics, University of Wroclaw, Wroclaw, Poland}
\end{Authlist}

\section*{Collaboration Institutes}
\renewcommand\theenumi{\arabic{enumi}~}
\begin{Authlist}

\item \Idef{org36632}Academy of Scientific Research and Technology (ASRT), Cairo, Egypt
\item \Idef{org1332}A. I. Alikhanyan National Science Laboratory (Yerevan Physics Institute) Foundation, Yerevan, Armenia
\item \Idef{org1279}Benem\'{e}rita Universidad Aut\'{o}noma de Puebla, Puebla, Mexico
\item \Idef{org1220}Bogolyubov Institute for Theoretical Physics, Kiev, Ukraine
\item \Idef{org20959}Bose Institute, Department of Physics and Centre for Astroparticle Physics and Space Science (CAPSS), Kolkata, India
\item \Idef{org1262}Budker Institute for Nuclear Physics, Novosibirsk, Russia
\item \Idef{org1292}California Polytechnic State University, San Luis Obispo, California, United States
\item \Idef{org1329}Central China Normal University, Wuhan, China
\item \Idef{org14939}Centre de Calcul de l'IN2P3, Villeurbanne, France
\item \Idef{org1197}Centro de Aplicaciones Tecnol\'{o}gicas y Desarrollo Nuclear (CEADEN), Havana, Cuba
\item \Idef{org1242}Centro de Investigaciones Energ\'{e}ticas Medioambientales y Tecnol\'{o}gicas (CIEMAT), Madrid, Spain
\item \Idef{org1244}Centro de Investigaci\'{o}n y de Estudios Avanzados (CINVESTAV), Mexico City and M\'{e}rida, Mexico
\item \Idef{org1335}Centro Fermi - Museo Storico della Fisica e Centro Studi e Ricerche ``Enrico Fermi'', Rome, Italy
\item \Idef{org17347}Chicago State University, Chicago, United States
\item \Idef{org1288}Commissariat \`{a} l'Energie Atomique, IRFU, Saclay, France
\item \Idef{org15782}COMSATS Institute of Information Technology (CIIT), Islamabad, Pakistan
\item \Idef{org1294}Departamento de F\'{\i}sica de Part\'{\i}culas and IGFAE, Universidad de Santiago de Compostela, Santiago de Compostela, Spain
\item \Idef{org1106}Department of Physics Aligarh Muslim University, Aligarh, India
\item \Idef{org1121}Department of Physics and Technology, University of Bergen, Bergen, Norway
\item \Idef{org1162}Department of Physics, Ohio State University, Columbus, Ohio, United States
\item \Idef{org1300}Department of Physics, Sejong University, Seoul, South Korea
\item \Idef{org1268}Department of Physics, University of Oslo, Oslo, Norway
\item \Idef{org1315}Dipartimento di Fisica dell'Universit\`{a} and Sezione INFN, Trieste, Italy
\item \Idef{org1145}Dipartimento di Fisica dell'Universit\`{a} and Sezione INFN, Cagliari, Italy
\item \Idef{org1312}Dipartimento di Fisica dell'Universit\`{a} and Sezione INFN, Turin, Italy
\item \Idef{org1285}Dipartimento di Fisica dell'Universit\`{a} `La Sapienza' and Sezione INFN, Rome, Italy
\item \Idef{org1154}Dipartimento di Fisica e Astronomia dell'Universit\`{a} and Sezione INFN, Catania, Italy
\item \Idef{org1132}Dipartimento di Fisica e Astronomia dell'Universit\`{a} and Sezione INFN, Bologna, Italy
\item \Idef{org1270}Dipartimento di Fisica e Astronomia dell'Universit\`{a} and Sezione INFN, Padova, Italy
\item \Idef{org1290}Dipartimento di Fisica `E.R.~Caianiello' dell'Universit\`{a} and Gruppo Collegato INFN, Salerno, Italy
\item \Idef{org1103}Dipartimento di Scienze e Innovazione Tecnologica dell'Universit\`{a} del Piemonte Orientale and Gruppo Collegato INFN, Alessandria, Italy
\item \Idef{org1114}Dipartimento Interateneo di Fisica `M.~Merlin' and Sezione INFN, Bari, Italy
\item \Idef{org1237}Division of Experimental High Energy Physics, University of Lund, Lund, Sweden
\item \Idef{org1192}European Organization for Nuclear Research (CERN), Geneva, Switzerland
\item \Idef{org1227}Fachhochschule K\"{o}ln, K\"{o}ln, Germany
\item \Idef{org1122}Faculty of Engineering, Bergen University College, Bergen, Norway
\item \Idef{org1136}Faculty of Mathematics, Physics and Informatics, Comenius University, Bratislava, Slovakia
\item \Idef{org1274}Faculty of Nuclear Sciences and Physical Engineering, Czech Technical University in Prague, Prague, Czech Republic
\item \Idef{org1229}Faculty of Science, P.J.~\v{S}af\'{a}rik University, Ko\v{s}ice, Slovakia
\item \Idef{org1184}Frankfurt Institute for Advanced Studies, Johann Wolfgang Goethe-Universit\"{a}t Frankfurt, Frankfurt, Germany
\item \Idef{org1215}Gangneung-Wonju National University, Gangneung, South Korea
\item \Idef{org20958}Gauhati University, Department of Physics, Guwahati, India
\item \Idef{org1212}Helsinki Institute of Physics (HIP) and University of Jyv\"{a}skyl\"{a}, Jyv\"{a}skyl\"{a}, Finland
\item \Idef{org1203}Hiroshima University, Hiroshima, Japan
\item \Idef{org1254}Indian Institute of Technology Bombay (IIT), Mumbai, India
\item \Idef{org36378}Indian Institute of Technology Indore, Indore, India (IITI)
\item \Idef{org1266}Institut de Physique Nucl\'{e}aire d'Orsay (IPNO), Universit\'{e} Paris-Sud, CNRS-IN2P3, Orsay, France
\item \Idef{org1277}Institute for High Energy Physics, Protvino, Russia
\item \Idef{org1249}Institute for Nuclear Research, Academy of Sciences, Moscow, Russia
\item \Idef{org1320}Nikhef, National Institute for Subatomic Physics and Institute for Subatomic Physics of Utrecht University, Utrecht, Netherlands
\item \Idef{org1250}Institute for Theoretical and Experimental Physics, Moscow, Russia
\item \Idef{org1230}Institute of Experimental Physics, Slovak Academy of Sciences, Ko\v{s}ice, Slovakia
\item \Idef{org1127}Institute of Physics, Bhubaneswar, India
\item \Idef{org1275}Institute of Physics, Academy of Sciences of the Czech Republic, Prague, Czech Republic
\item \Idef{org1139}Institute of Space Sciences (ISS), Bucharest, Romania
\item \Idef{org27399}Institut f\"{u}r Informatik, Johann Wolfgang Goethe-Universit\"{a}t Frankfurt, Frankfurt, Germany
\item \Idef{org1185}Institut f\"{u}r Kernphysik, Johann Wolfgang Goethe-Universit\"{a}t Frankfurt, Frankfurt, Germany
\item \Idef{org1177}Institut f\"{u}r Kernphysik, Technische Universit\"{a}t Darmstadt, Darmstadt, Germany
\item \Idef{org1256}Institut f\"{u}r Kernphysik, Westf\"{a}lische Wilhelms-Universit\"{a}t M\"{u}nster, M\"{u}nster, Germany
\item \Idef{org1246}Instituto de Ciencias Nucleares, Universidad Nacional Aut\'{o}noma de M\'{e}xico, Mexico City, Mexico
\item \Idef{org1247}Instituto de F\'{\i}sica, Universidad Nacional Aut\'{o}noma de M\'{e}xico, Mexico City, Mexico
\item \Idef{org1308}Institut Pluridisciplinaire Hubert Curien (IPHC), Universit\'{e} de Strasbourg, CNRS-IN2P3, Strasbourg, France
\item \Idef{org1182}Joint Institute for Nuclear Research (JINR), Dubna, Russia
\item \Idef{org1199}Kirchhoff-Institut f\"{u}r Physik, Ruprecht-Karls-Universit\"{a}t Heidelberg, Heidelberg, Germany
\item \Idef{org20954}Korea Institute of Science and Technology Information, Daejeon, South Korea
\item \Idef{org1017642}KTO Karatay University, Konya, Turkey
\item \Idef{org1160}Laboratoire de Physique Corpusculaire (LPC), Clermont Universit\'{e}, Universit\'{e} Blaise Pascal, CNRS--IN2P3, Clermont-Ferrand, France
\item \Idef{org1194}Laboratoire de Physique Subatomique et de Cosmologie (LPSC), Universit\'{e} Joseph Fourier, CNRS-IN2P3, Institut Polytechnique de Grenoble, Grenoble, France
\item \Idef{org1187}Laboratori Nazionali di Frascati, INFN, Frascati, Italy
\item \Idef{org1232}Laboratori Nazionali di Legnaro, INFN, Legnaro, Italy
\item \Idef{org1125}Lawrence Berkeley National Laboratory, Berkeley, California, United States
\item \Idef{org1234}Lawrence Livermore National Laboratory, Livermore, California, United States
\item \Idef{org1251}Moscow Engineering Physics Institute, Moscow, Russia
\item \Idef{org1322}National Centre for Nuclear Studies, Warsaw, Poland
\item \Idef{org1140}National Institute for Physics and Nuclear Engineering, Bucharest, Romania
\item \Idef{org1017626}National Institute of Science Education and Research, Bhubaneswar, India
\item \Idef{org1165}Niels Bohr Institute, University of Copenhagen, Copenhagen, Denmark
\item \Idef{org1109}Nikhef, National Institute for Subatomic Physics, Amsterdam, Netherlands
\item \Idef{org1283}Nuclear Physics Institute, Academy of Sciences of the Czech Republic, \v{R}e\v{z} u Prahy, Czech Republic
\item \Idef{org1264}Oak Ridge National Laboratory, Oak Ridge, Tennessee, United States
\item \Idef{org1189}Petersburg Nuclear Physics Institute, Gatchina, Russia
\item \Idef{org1170}Physics Department, Creighton University, Omaha, Nebraska, United States
\item \Idef{org1157}Physics Department, Panjab University, Chandigarh, India
\item \Idef{org1112}Physics Department, University of Athens, Athens, Greece
\item \Idef{org1152}Physics Department, University of Cape Town and  iThemba LABS, National Research Foundation, Somerset West, South Africa
\item \Idef{org1209}Physics Department, University of Jammu, Jammu, India
\item \Idef{org1207}Physics Department, University of Rajasthan, Jaipur, India
\item \Idef{org1200}Physikalisches Institut, Ruprecht-Karls-Universit\"{a}t Heidelberg, Heidelberg, Germany
\item \Idef{org1017688}Politecnico di Torino, Turin, Italy
\item \Idef{org1325}Purdue University, West Lafayette, Indiana, United States
\item \Idef{org1281}Pusan National University, Pusan, South Korea
\item \Idef{org1176}Research Division and ExtreMe Matter Institute EMMI, GSI Helmholtzzentrum f\"ur Schwerionenforschung, Darmstadt, Germany
\item \Idef{org1334}Rudjer Bo\v{s}kovi\'{c} Institute, Zagreb, Croatia
\item \Idef{org1298}Russian Federal Nuclear Center (VNIIEF), Sarov, Russia
\item \Idef{org1252}Russian Research Centre Kurchatov Institute, Moscow, Russia
\item \Idef{org1224}Saha Institute of Nuclear Physics, Kolkata, India
\item \Idef{org1130}School of Physics and Astronomy, University of Birmingham, Birmingham, United Kingdom
\item \Idef{org1338}Secci\'{o}n F\'{\i}sica, Departamento de Ciencias, Pontificia Universidad Cat\'{o}lica del Per\'{u}, Lima, Peru
\item \Idef{org1155}Sezione INFN, Catania, Italy
\item \Idef{org1313}Sezione INFN, Turin, Italy
\item \Idef{org1271}Sezione INFN, Padova, Italy
\item \Idef{org1133}Sezione INFN, Bologna, Italy
\item \Idef{org1146}Sezione INFN, Cagliari, Italy
\item \Idef{org1316}Sezione INFN, Trieste, Italy
\item \Idef{org1115}Sezione INFN, Bari, Italy
\item \Idef{org1286}Sezione INFN, Rome, Italy
\item \Idef{org36377}Nuclear Physics Group, STFC Daresbury Laboratory, Daresbury, United Kingdom
\item \Idef{org1258}SUBATECH, Ecole des Mines de Nantes, Universit\'{e} de Nantes, CNRS-IN2P3, Nantes, France
\item \Idef{org35706}Suranaree University of Technology, Nakhon Ratchasima, Thailand
\item \Idef{org1304}Technical University of Split FESB, Split, Croatia
\item \Idef{org1017659}Technische Universit\"{a}t M\"{u}nchen, Munich, Germany
\item \Idef{org1168}The Henryk Niewodniczanski Institute of Nuclear Physics, Polish Academy of Sciences, Cracow, Poland
\item \Idef{org17361}The University of Texas at Austin, Physics Department, Austin, TX, United States
\item \Idef{org1173}Universidad Aut\'{o}noma de Sinaloa, Culiac\'{a}n, Mexico
\item \Idef{org1296}Universidade de S\~{a}o Paulo (USP), S\~{a}o Paulo, Brazil
\item \Idef{org1149}Universidade Estadual de Campinas (UNICAMP), Campinas, Brazil
\item \Idef{org1239}Universit\'{e} de Lyon, Universit\'{e} Lyon 1, CNRS/IN2P3, IPN-Lyon, Villeurbanne, France
\item \Idef{org1205}University of Houston, Houston, Texas, United States
\item \Idef{org20371}University of Technology and Austrian Academy of Sciences, Vienna, Austria
\item \Idef{org1222}University of Tennessee, Knoxville, Tennessee, United States
\item \Idef{org1310}University of Tokyo, Tokyo, Japan
\item \Idef{org1318}University of Tsukuba, Tsukuba, Japan
\item \Idef{org21360}Eberhard Karls Universit\"{a}t T\"{u}bingen, T\"{u}bingen, Germany
\item \Idef{org1225}Variable Energy Cyclotron Centre, Kolkata, India
\item \Idef{org1017687}Vestfold University College, Tonsberg, Norway
\item \Idef{org1306}V.~Fock Institute for Physics, St. Petersburg State University, St. Petersburg, Russia
\item \Idef{org1323}Warsaw University of Technology, Warsaw, Poland
\item \Idef{org1179}Wayne State University, Detroit, Michigan, United States
\item \Idef{org1143}Wigner Research Centre for Physics, Hungarian Academy of Sciences, Budapest, Hungary
\item \Idef{org1260}Yale University, New Haven, Connecticut, United States
\item \Idef{org15649}Yildiz Technical University, Istanbul, Turkey
\item \Idef{org1301}Yonsei University, Seoul, South Korea
\item \Idef{org1327}Zentrum f\"{u}r Technologietransfer und Telekommunikation (ZTT), Fachhochschule Worms, Worms, Germany
\end{Authlist}
\endgroup

\end{document}